# The dissociation catastrophe in fluctuating-charge models and its implications for the concept of atomic electronegativity

**Jiahao Chen[1] and Todd J. Martínez[1]**

[1]Department of Chemistry, Center for Advanced Theory and Molecular Simulation, Frederick Seitz Materials Research Laboratory, and The Beckman Institute

University of Illinois at Urbana–Champaign, Urbana, Illinois 61801

**Abstract**  We have recently developed the QTPIE (charge transfer with polarization current equilibration) fluctuating–charge model, a new model with correct dissociation behavior for nonequilibrium geometries. The correct asymptotics originally came at the price of representing the solution in terms of charge–transfer variables instead of atomic charges. However, we have found an exact reformulation of fluctuating–charge models in terms of atomic charges again, which is made possible by the symmetries of classical electrostatics. We show how this leads to the distinguishing between two types of atomic electronegativities in our model. While one is a intrinsic property of individual atoms, the other takes into account the local electrical surroundings. This suggests that this distinction could resolve some confusion surrounding the concept of electronegativity as to whether it is an intrinsic property of elements, or otherwise.

## Introduction

Recent studies using classical molecular dynamics have found conventional additive force fields increasingly inadequate for today's systems of interest, as the neglect of nonadditive phenomena such as polarization and charge transfer can lead to qualitative changes in simulations.[1-4] Of the two nonadditive effects, the literature on methods to incorporate polarization is more extensive. Two of the many popular types of methods for incorporating polarization are inducible di-poles[3-5], where additional variables are introduced to describe dipole moments induced by mutual polarization interactions; and Drude oscillators[6, 7], where polarization is described by the change in distance between the atomic nucleus and a fixed countercharge attached by a harmonic potential. However, neither of these methods are readily extensible to provide a description of charge transfer. This is in some sense surprising, as charge transfer is merely an extreme form of polarization: while polarization results in a redistribution of charge density within molecules, charge transfer is a redistribution of charge density across molecules.

In contrast, there are several classes of methods that exist for modeling both charge transfer and polarization effects: for example, fluctuating–charge models,[2, 8, 9] which model polarization by recomputing the charge distribution in response to changes in geometry or external perturbations; empirical valence bond (EVB) methods,[10-12] which parameterize the energy contributions of individual valence bond configurations; and effective fragment potential (EFP)–type methods,[13, 14] which use energy decompositions of *ab initio* data to construct parameterized effective potentials.

We choose to study only fluctuating–charge models, as the other methods that treat both polarization and charge transfer are computationally far more costly. In EFPs, polarization is modeled using distributed, inducible dipoles while charge transfer is represented separately as a sum over antibonding orbitals of the electron acceptor. The latter necessitates *a priori* specification of the charge acceptors and donors, as well as the provision of parameters for every orbital being summed over. Not only is this description computationally expensive, but it also fails to provide a unified picture of polarization and charge transfer. In contrast, EVB does provide this unified treatment, but suffers from the exponential growth in the number of relevant valence bond configurations with system size. In contrast, fluctuating–charge models introduce only a modest computational cost over conventional fixed–charge force fields, even for large systems. Several of these methods have been used in dynamics simulations, most notably QEq[15] in UFF[16], EEM[17, 18] in ReaxFF[19] and *fluc*-q in the TIP4P-FQ water model[20, 21], thus demonstrating their utility in describing polarization effects in classical molecular dynamics.

In addition, fluctuating–charge models are theoretically appealing as they provide a unified treatment of polarization and charge transfer with only two parameters per atom. These parameters can be identified with the chemically important concepts of electronegativity[22-28] and (chemical) hardness[29-32]. These drive the redistribution of atomic charges in response to electrostatic interactions according to the principle of electronegativity equalization.[24-27, 33]

# The dissociation catastrophe in QEq-type fluctuating-charge models

Here, we briefly review the most common type of fluctuating-charge model and how such models are solved. The main idea of fluctuating-charge models is to assert that the electrostatic energy of a molecular system can be decomposed into two types of terms, i.e.

$$E(\mathbf{q};\mathbf{R}) = \sum_{i=1}^{N} E_i^{at}(q_i) + \sum_{i<j} q_i q_j J_{ij}(\mathbf{R}) \qquad (1)$$

where $N$ is the number of atoms in the system, $\mathbf{q} = (q_1,...,q_N)$ are the charges on each atom, each $E_i^{at}$ is the intrinsic contribution of each individual atom, and each $J_{ij}$ is a pairwise interaction that is dependent on the molecular geometry $\mathbf{R} = (\mathbf{R}_1,...,\mathbf{R}_N)$. The atomic charges are then solved for by a minimization of the total electrostatic energy with respect to each atomic charge with a constraint on the total charge of the system, $Q$:

$$\sum_{i=1}^{N} q_i = Q \qquad (2)$$

In many fluctuating-charge models, the interactions $J_{ij}$ are taken to represent some screened Coulomb interactions $J_{ij} = J_{ij}(|\mathbf{R}_i - \mathbf{R}_j|)$ that do not diverge in the small separation limit $|\mathbf{R}_i - \mathbf{R}_j| \rightarrow 0$. Screening is necessary in order to prevent numerical instabilities from occurring at small interatomic distances.

The precise method of calculating these interactions differs between the specific fluctuating–charge models in the literature: in the electronegativity equalization method (EEM)[17, 18], the Coulomb interactions are evaluated as two-electron Coulomb integrals over spherically symmetric Gaussian-type atomic orbitals; the chemical potential equalization (CPE)[34] model uses similar integrals, but with empirical parameters for Fukui function corrections; in the QEq[15] and fluc-$q$[20, 21] models, the Coulomb interactions are evaluated as two-electron Coulomb integrals over spherically symmetric Slater-type atomic orbitals; and in the CHARMM C22 force field,[35, 36] the Coulomb interactions are screening with empirical functions. In the QTPIE model,[37] we use two–electron Coulomb integrals over $s$–type primitive Gaussian orbitals, with orbital exponents fitted to reproduce the results from the much more expensive $s$–type Slater type orbitals used in QEq. We have found that it is possible to optimize Gaussian orbitals to reproduce the Slater integrals with an accuracy of less than $10^{-3}$ atomic units, with exponents given in Table 1. The details of the fitting procedure are given in Appendix A.

In addition, the atomic terms $E_i^{at}\left(q_i\right)$ in many fluctuating-charge models are each assumed to be a quadratic polynomial of the form

$$E_i^{at}\left(q_i\right) = E_i^0 + \chi_i q_i + \tfrac{1}{2}\eta_i q_i^2 + \ldots \tag{3}$$

where $E_i^0$ is a constant independent of charge and geometry and can thus be discarded in the energy expression for fluctuating–charge models. The other coefficients are interpreted by a formal comparison with a Taylor series expansion of $E_i^{at}\left(q_i\right)$ about $q_i = 0$.[24]

$$E_i^{at}\left(q_i\right) = E_i^{at}\left(0\right) + \left.\frac{dE_i^{at}}{dq_i}\right|_{q_i=0}\left(q_i - 0\right) + \frac{1}{2!}\left.\frac{d^2 E_i^{at}}{dq_i^2}\right|_{q_i=0}\left(q_i - 0\right)^2 + \ldots \tag{4}$$

By approximating the Taylor expansion coefficients with suitable finite difference formulas with spacing $\Delta q_i = 1$, the following well—known relationships are obtained:

$$\chi_i \equiv \left.\frac{dE_i^{at}}{dq_i}\right|_{q_i=0} \approx \frac{E_i^{at}\left(1\right) - E_i^{at}\left(-1\right)}{2} = \frac{\mathrm{IP}_i + \mathrm{EA}_i}{2} \tag{5}$$

$$\eta_i \equiv \left.\frac{d^2 E_i^{at}}{dq_i^2}\right|_{q_i=0} \approx E_i^{at}\left(1\right) - 2E_i^{at}\left(0\right) + E_i^{at}\left(-1\right) = \mathrm{IP}_i - \mathrm{EA}_i \tag{6}$$

where $\mathrm{IP}_i = E_i^{at}\left(1\right) - E_i^{at}\left(0\right)$ is the ionization potential of the $i^{\mathrm{th}}$ atom and $\mathrm{EA}_i = E_i^{at}\left(0\right) - E_i^{at}\left(-1\right)$ is the electron affinity of the $i^{\mathrm{th}}$ atom. In this manner, these coefficients can be identified as none other than the Mulliken electronegativity[38] and Parr-Pearson (chemical) hardness[32] respectively. The preceding identifications allow fluctuating-charge models to be identified as rudimentary forms of density functional theory.[39]

The truncation of the series expansion (4) at second order allows the solution to be found by solving a linear system of equations. The only complication is the need to enforce the constraint (2), which can be taken care of with the method of Lagrange multipliers. In this context, the Lagrange multiplier $\mu$ can be interpreted as the chemical potential, and the solution to the constrained problem is the charge distribution and chemical potential which minimizes the free energy

$$F(\mathbf{q}, \mu; Q) = E(\mathbf{q}) - \mu\left(\sum_{i=1}^{N} q_i - Q\right)$$

$$= \mu Q + \sum_{i=1}^{N}(\chi_i - \mu)q_i + \tfrac{1}{2}\sum_{ij} q_i q_j J_{ij}$$

(7)

where $\eta_i = J_{ii}$. Minimizing this free energy then leads to the linear system of equations consisting of (2) and the equations

$$0 = \frac{\partial F(\mathbf{q})}{\partial q_i} = (\chi_i - \mu) + \sum_{j=1}^{N} q_j J_{ij}$$

(8)

This system can be written in block-matrix notation

$$\begin{pmatrix} \mathbf{J} & \mathbf{1} \\ \mathbf{1}^T & 0 \end{pmatrix} \begin{pmatrix} \mathbf{q} \\ \mu \end{pmatrix} = \begin{pmatrix} -\boldsymbol{\chi} \\ Q \end{pmatrix}$$

(9)

where $\mathbf{1}$ is a column vector with entries all equal to unity. This system of equations is solved approximately in the historically important models of Del Re[40] and Gasteiger and Marsili[41]; however, all modern models solve these equations exactly for the charge distribution. It is straightforward to show (as in Appendix B) that this linear system has the explicit solution

$$\begin{pmatrix} \mathbf{q} \\ \mu \end{pmatrix} = \begin{pmatrix} -\mathbf{J}^{-1}(\boldsymbol{\chi} + \mu\mathbf{1}) \\ -(Q + \mathbf{1}^T\mathbf{J}^{-1}\boldsymbol{\chi})/\mathbf{1}^T\mathbf{J}^{-1}\mathbf{1} \end{pmatrix}$$

(10)

It is instructive to solve the fluctuating-charge model above in the case of a neutral diatomic molecule. Then, (7) can be written explicitly in terms of one charge variable $q_1$, so that the energy is given by

$$F(q_1; \mathbf{R}) = (\chi_1 - \chi_2)q_1 + \tfrac{1}{2}\left(\eta_1 - 2J_{12}\left(|\mathbf{R}_1 - \mathbf{R}_2|\right) + \eta_2\right)q_1^2$$

(11)

This is minimized by the explicit solution

$$q_1(\mathbf{R}) = \frac{\chi_2 - \chi_1}{\eta_1 - 2J_{12}\left(|\mathbf{R}_1 - \mathbf{R}_2|\right) + \eta_2}$$

(12)

We therefore see that this fluctuating-charge model always predicts a nonzero charge on each atom unless they have equal electronegativities or at least one atom has infinite hardness. While this is reasonable for chemically bonded sys-tems, it fails to describe, even qualitatively, the charge transfer behavior at infinite separation. As $\left|\mathbf{R}_1 - \mathbf{R}_2\right| \to \infty$, the Coulomb interaction vanishes, so that

$$\lim_{|\mathbf{R}_1 - \mathbf{R}_2| \to \infty} q_1(\mathbf{R}) = \frac{\chi_2 - \chi_1}{\eta_1 + \eta_2} \neq 0 \tag{13}$$

The model therefore predicts nonzero charge transfer even for dissociated systems, which is clearly unphysical for diatomic molecules in the gas phase. This leads to a dissociation catastrophe whereby intermolecular charge transfer is severely overestimated, causing electrostatic properties such as the dipole moment and the on-axis component of the polarizability to diverge. This renders such models useless for describing intermolecular charge transfers, and in addition requires further constraints proscribing intermolecular charge transfer in practical simulations.

This unphysical prediction of nonzero charge transfer at infinity can be understood by turning off the Coulomb interaction terms in (1). Then, the noninteracting energy $E^{\mathrm{NI}}$ becomes the simple sum

$$E^{\mathrm{NI}}(\mathbf{q}; \mathbf{R}) = \sum_{i=1}^{N} E_i^{at}(q_i) \tag{14}$$

which for the case of quadratic atomic energies (3) can be written in the form

$$E_i^{at}(q_i) = \tfrac{1}{2} \eta_i \left( q_i + \frac{\chi_i}{\eta_i} \right)^2 - \frac{\chi_i^2}{2\eta_i} + E_i^0 \tag{15}$$

Thus in the absence of any interatomic interactions, the charge predicted by fluctuating–charge models defaults to the solution $q_i = -\chi_i / \eta_i$, being the minimum point of the parabola (15). As both the atomic electronegativity and atomic hardness are constants, it is unclear how this problem can be solved while remaining in atom space, i.e. the solution space spanned by the vector of atomic charges $\mathbf{q}$.

The dissociation catastrophe can be interpreted as the consequence of an unrealistic assumption inherent in fluctuating-charge models, namely that pairs of atoms can exchange charge with equal facility regardless of their distance. This is true only in metallic phases, and therefore the extent to which this model fails to predict sensible charge distributions can be attributed to a fault in the underlying physics in assuming that molecular systems have metallic character. In the next section, we will discuss how to undo this assumption.

## The QTPIE model

In order to address this dissociation catastrophe, we have proposed the QTPIE (charge transfer with polarization current equilibration) model,[37, 42] which was first formulated not in terms of atomic charges, but in terms of charge-transfer variables,[43, 44] sometimes called split-charge variables.[45] These new variables $p_{ji}$ define a new solution space which we call the bond-space, and they account for the amount of charge that has flowed from the $j^{th}$ atom to the $i^{th}$ atom, and can be interpreted as the integral of a transient current between these two atoms. We require these variables to be antisymmetric, so that $p_{ji} = -p_{ij}$. Furthermore, we recover the atomic charges by summing over all source atoms, and thus the charge–transfer variables are related to the charge variables by the relation

$$\sum_{j=1}^{N} p_{ji} = q_i \qquad (16)$$

By applying this relation, the energy function of the QEq-type fluctuating-charge model (3) can be rewritten in terms of charge-transfer variables as

$$E(\mathbf{p}) = \sum_{i,j=1}^{N} \chi_i p_{ji} + \tfrac{1}{2} \sum_{i,j,k,l=1}^{N} p_{ki} p_{lj} J_{ij}$$
$$= \sum_{i<j} (\chi_i - \chi_j) p_{ji} + \tfrac{1}{2} \sum_{i<j,k<l} p_{ki} p_{lj} (J_{ij} - J_{il} - J_{kj} + J_{kl}) \qquad (17)$$

where on the second line, we have exploited the skew-symmetry of the charge-transfer variables. We note that the constraint (16) already enforces overall charge neutrality, i.e. $Q = 0$, and therefore the model can be solved immediately by direct minimization of this energy with respect to the charge-transfer variables $\mathbf{p}$ without the use of the Lagrange multipliers.

We now create the QTPIE model by modifying the first term in (17) to have a pairwise and geometry-dependent electronegativity. By replacing $\chi_i$ with $\tilde{\chi}_{ji}(\mathbf{R})$, we have the *new* energy function

$$E^{QTPIE}(\mathbf{p};\mathbf{R}) = \sum_{i,j=1}^{N} \tilde{\chi}_{ji}(\mathbf{R}) p_{ji} + \tfrac{1}{2} \sum_{i,j,k,l=1}^{N} p_{ki} p_{lj} J_{ij} (|\mathbf{R}_i - \mathbf{R}_j|)$$
$$= \sum_{i<j} (\tilde{\chi}_{ji} - \tilde{\chi}_{ij}) p_{ji} + \tfrac{1}{2} \sum_{i<j,k<l} p_{ki} p_{lj} (J_{ij} - J_{il} - J_{kj} + J_{kl})$$
$$(18)$$

Equation (18) defines the QTPIE model, which is solved in bond space by the solution to the linear system of equations

$$\tilde{\mathbf{J}}\mathbf{p} = \tilde{\mathbf{v}} \tag{19}$$

where the collection of charge–transfer variables $p_{(j,i)} = p_{ji}$ is now interpreted as a vector indexed by the multi-index $(j,i), 1 \leq i < j \leq N$. This defines a new vector space, which we call the bond space, with the bond hardness matrix $\tilde{J}_{(k,i),(l,j)} = J_{ij} - J_{il} - J_{kj} + J_{kl}$, and bond electronegativities $\tilde{v}_{(j,i)} = \tilde{\chi}_{ij} - \tilde{\chi}_{ji}$ that correspond to pairwise voltage differences.

For comparison purposes, we solve the QTPIE model analytically for a diatomic molecule. The model then consists of only one unknown variable $p_{21}$, and the model predicts the energy function

$$\begin{aligned} E^{QTPIE}\left(p_{21}; \mathbf{R}\right) &= \left(\tilde{\chi}_{21}(\mathbf{R}) - \tilde{\chi}_{12}(\mathbf{R})\right) p_{21} \\ &\quad + \tfrac{1}{2}\left(J_{11} - 2J_{12}\left(\left|\mathbf{R}_1 - \mathbf{R}_2\right|\right) + J_{22}\right) p_{21}^2 \end{aligned} \tag{20}$$

which has the solution

$$q_1(\mathbf{R}) = p_{21}(\mathbf{R}) = \frac{\tilde{\chi}_{12}(\mathbf{R}) - \tilde{\chi}_{21}(\mathbf{R})}{J_{11} - 2J_{12}\left(\left|\mathbf{R}_1 - \mathbf{R}_2\right|\right) + J_{22}} \tag{21}$$

In contrast to (12), it is possible to attenuate long-distance charge transfer as $\left|\mathbf{R}_1 - \mathbf{R}_2\right| \rightarrow \infty$ by requiring that $\tilde{\chi}_{12}(\mathbf{R}) - \tilde{\chi}_{21}(\mathbf{R}) \rightarrow 0$ at the same time. In the QTPIE model, there are several reasonable choices for the pairwise electronegativity,[42] but we believe a reasonable definition of the pairwise electronegativity is

$$\tilde{\chi}_{ij}(\mathbf{R}) - \tilde{\chi}_{ji}(\mathbf{R}) = \frac{S_{ij}\left(\left|\mathbf{R}_i - \mathbf{R}_j\right|\right)}{\left\langle S_{ij'}\left(\left|\mathbf{R}_i - \mathbf{R}_{j'}\right|\right)\right\rangle_{j'}}\left(\chi_i - \chi_j\right) \tag{22}$$

which is essentially the bare atomic electronegativity $\chi_i$ weighted by $S_{ij}\left(\left|\mathbf{R}_i - \mathbf{R}_j\right|\right) = \int \phi_i\left(\mathbf{r}_1; \mathbf{R}_i\right)\phi_j\left(\mathbf{r}_1; \mathbf{R}_j\right)d\mathbf{r}_1$, the overlap integral between the atomic orbitals on the $i^{\text{th}}$ and $j^{\text{th}}$ atoms as introduced to calculate the screened Coulomb interactions ( 33) as described in Appendix A, and renormalized by a system-dependent constant that rescales the weighting factor by the *average* weighting factor over all atoms, $\left\langle S_{i'j}\left(\left|\mathbf{R}_{i'} - \mathbf{R}_j\right|\right)\right\rangle_{i'} = \sum_{i'=1}^{N} S_{i'j}\left(\left|\mathbf{R}_{i'} - \mathbf{R}_j\right|\right) / N$.

Indeed, this choice of pairwise electronegativity produces the correct asymptotic limit of no charge transfer at infinite separation as the electronegativity difference vanishes due to the asymptotic property of the overlap integral that $S_{ij} \to 0$ as $\left| \mathbf{R}_i - \mathbf{R}_j \right| \to \infty$. Note that if we set all the attenuation factors to a numerical constant, say $S_{ij} = 1$, (22) reduces to just $\chi_i$ and we recover the QEq-type fluctuating-charge model of the preceding section. The dissociation catastrophe returns when the model assumes that pairs of atoms can exchange charge with equal facility regardless of their distance, thus reinforcing our earlier observation that the failures of the earlier model can be attributed to assuming that all systems to have metallic electronic structure.

## The exact reformulation of models in bond space as models in atom space

The preceding discussion shows that we have found a solution to the dissociation catastrophe, and therefore have a framework for fluctuating–charge models that are useful for describing intermolecular charge transfer. However, this apparently comes at the price of representing the solution in bond space, which for a $N$–atom system has $N$ times as many variables as the original representation in atom space. Perhaps surprisingly, it is possible to reformulate an arbitrary fluctuating–charge model formulated in bond space exactly as an equivalent fluctuating–charge model in atom space.

The key insight is that the bond–space hardness matrix $\tilde{\mathbf{J}}$ in (19) is rank deficient, and that its nullspace is spanned by vectors describing cyclic charge transport.[42] In order to show this, we note that relationship between charges and charge-transfer variables (16) is linear. Therefore the mapping from solutions in bond space to those in atom space can be expressed by a rectangular matrix $\mathbf{T}$ such that

$$\mathbf{T} : \mathbb{R}^{N(N-1)/2} \to \mathbb{R}^{N-1} \cong \left\{ \mathbf{q} \in \mathbb{R}^N \mid \mathbf{1}^T \mathbf{q} = 0 \right\}$$

$$\mathbf{T}\mathbf{p} = \mathbf{q} \tag{23}$$

Then the relationship between $\mathbf{J}$ and $\tilde{\mathbf{J}}$ can be expressed as

$$\tilde{\mathbf{J}} = \mathbf{T}^T \mathbf{J} \mathbf{T} \tag{24}$$

We had previously introduced a directed graph $G$ whose vertices are in one–to–one correspondence with atomic charges and edges that are in one–to–one correspondence with charge–transfer variables. For fluctuating–charge models, $G$ is a complete graph. Then $\mathbf{T}$ corresponds to the adjacency matrix for $G$, which has ma-trix element $T_{ve}$ equal to 1 if the edge $e$ points toward the vertex $v$, $-1$ if the edge $e$ points away from the vertex $v$, and 0 otherwise.

While the atom–space hardness matrix $\mathbf{J}$ is of full rank, $\tilde{\mathbf{J}}$ has dimension $N(N-1)/2$ but only rank $N-1$. This is because combinations of charge transfer variables that correspond to cyclic charge transport belong to the null-space of $\tilde{\mathbf{J}}$. For illustrative purposes, consider a four–charge system that is described by the variables $\{q_1, q_2, q_3, q_4\}$ in atom space and $\{p_{21}, p_{31}, p_{41}, p_{32}, p_{42}, p_{43}\}$ in bond space. For this system, $\mathbf{T}$ has the representation

$$\mathbf{T} = \begin{pmatrix} 1 & 1 & 1 & 0 & 0 & 0 \\ -1 & 0 & 0 & 1 & 1 & 0 \\ 0 & -1 & 0 & -1 & 0 & 1 \\ 0 & 0 & -1 & 0 & -1 & -1 \end{pmatrix} \qquad (25)$$

Consider the combination of charge transfer variables $p_{12} + p_{23} + p_{34} + p_{41} = -p_{21} + p_{41} - p_{32} - p_{43}$. In the current basis, this corresponds to the vector $\boldsymbol{\Gamma} = (-1, 0, 1, -1, 0, -1)^T$. Then it is straightforward to verify by explicit calculation that

$$\mathbf{T}\boldsymbol{\Gamma} = \begin{pmatrix} 1 & 1 & 1 & 0 & 0 & 0 \\ -1 & 0 & 0 & 1 & 1 & 0 \\ 0 & -1 & 0 & -1 & 0 & 1 \\ 0 & 0 & -1 & 0 & -1 & -1 \end{pmatrix} \begin{pmatrix} -1 \\ 0 \\ 1 \\ -1 \\ 0 \\ -1 \end{pmatrix} = \begin{pmatrix} 0 \\ 0 \\ 0 \\ 0 \end{pmatrix} = \mathbf{0} \quad (26)$$

Hence $\boldsymbol{\Gamma}$, which represents a cyclic flow of charge, is in the nullspace of $\mathbf{T}$, and hence by the relation ( 24 ) is also in the nullspace of $\tilde{\mathbf{J}}$. We interpret this as a consequence of Kirchhoff's voltage law, which arises from the conservative nature of the electrostatic potential. By similar calculations one can show that any cyclic combination of charge transfer variables lies in the nullspace of $\tilde{\mathbf{J}}$, and so acyclic combinations of charge transfer variables are the only ones that lie in the range of $\tilde{\mathbf{J}}$. An elementary result of graph theory immediately yields that the space of acyclic combinations of charge transfer variables is spanned by $N-1$ linearly independent vectors, and hence $\mathbf{T}$ is of rank $N-1$. We have previously provided a rigorous proof of this fact.[42]

When combined with the fact that $\mathbf{J}$ is a discretization of the Coulomb operator in a finite and localized basis, and is therefore full rank ($N$) for reasonable geometries that do not have degenerate or nearly coincident atoms, the composition ( 24) shows that $\tilde{\mathbf{J}}$ must have rank $N-1$, and that there are always $N-1$ physically important degrees of freedom, whether they are represented as atomic charges or charge transfer variables. This suggests that it is possible to reformulate exactly any fluctuating-charge model represented in bond space as an equivalent model in atom space. In order to do this, we require the inverse mapping $\mathbf{T}^+$ such that $\mathbf{p} = \mathbf{T}^+\mathbf{q}$. As $\mathbf{T}$ is rectangular, the conventional inverse $\mathbf{T}^{-1}$ cannot exist, but the preceding discussion shows that the generalized inverse such as the Moore-Penrose pseudoinverse $\mathbf{T}^+$ performs the same role as the conventional inverse in that no information is lost in the inversion. Interestingly, it is possible to verify by explicit calculation that for the complete graph $G$, $\mathbf{T}^+ = \mathbf{T}^T / N$, so that

$$p_{ba} = \frac{q_a - q_b}{N} \tag{27}$$

This simple relation allows the energy function of the QTPIE model (18) to be reformulated exactly as

$$
\begin{aligned}
E^{QTPIE}\left(\mathbf{T}^+\mathbf{q}\right) &= \tilde{\mathbf{v}}\mathbf{T}^+\mathbf{q} + \tfrac{1}{2}\mathbf{q}^T\mathbf{J}\mathbf{q} \\
&= \sum_{i=1}^{N} q_i \sum_{j=1}^{N} \frac{\tilde{\chi}_{ji}\left(\mathbf{R}\right) - \tilde{\chi}_{ij}\left(\mathbf{R}\right)}{N} + \frac{1}{2}\sum_{i,j=1}^{N} q_i q_j J_{ij}
\end{aligned}
\tag{28}
$$

Interestingly, this expression shows that the introduction of pairwise electronegativities results in an effective atomic electronegativity

$$\overline{\chi}_i = -\left(\mathbf{T}^+\mathbf{q}\right)_i = \sum_{j=1}^{N} \frac{\tilde{\chi}_{ij}\left(\mathbf{R}\right) - \tilde{\chi}_{ji}\left(\mathbf{R}\right)}{N} \tag{29}$$

which for the definition of the pairwise electronegativity (22) gives rise to

$$\overline{\chi}_i = \frac{\displaystyle\sum_{j=1}^{N} S_{ij}\left(\left|\mathbf{R}_i - \mathbf{R}_j\right|\right)\left(\chi_i - \chi_j\right)}{\displaystyle\sum_{j'=1}^{N} S_{ij'}\left(\left|\mathbf{R}_i - \mathbf{R}_{j'}\right|\right)} = \chi_i - \frac{\displaystyle\sum_{j=1}^{N} S_{ij}\left(\left|\mathbf{R}_i - \mathbf{R}_j\right|\right)\chi_j}{\displaystyle\sum_{j'=1}^{N} S_{ij'}\left(\left|\mathbf{R}_i - \mathbf{R}_{j'}\right|\right)} \tag{30}$$

that is in general different from the bare atomic electronegativity $\chi_i$ that goes into (22) due to the presence of an explicitly environment–dependent term. This is the main result of our paper.

## Application to simple hydrocarbons

In **Table 2**, we provide a simple illustration of the difference between the effective atomic electronegativity $\overline{\chi}_i$ (30) and the intrinsic atomic electronegativity $\chi_i$ in four simple hydrocarbon systems – staggered ethane, eclipsed ethane, ethylene and acetylene. For comparison purposes, we also provide electronegativities calculated from the Mulliken-Jaffé scale [46]. Furthermore, we contrast the charge distributions calculated in QTPIE, as well as in QEq, but without the adjustment of the hydrogen exponent as described in the original paper use to calculate a charge-dependent hardness, which we term QEq(-H). In addition, we calculated *ab initio* wavefunctions at the MP2/6-31G level of theory and perform two types of charge analysis, namely Mulliken population analysis[47] and zeroth-order distributed multipole analysis,[48] which we term distributed monopole analysis. Note that from the preceding discussion, there is no difference between the effective atomic electronegativity and the intrinsic atomic electronegativity in the QEq model. Furthermore, the energy function of fluctuating-charge models exhibits a global gauge symmetry in that the equations in atom space (7) and in bond space (17) are both invariant to a global shift in the intrinsic atomic electronegativities

$$\chi_i \mapsto \chi_i + V \tag{31}$$

where in the atom-space formulation, the global chemical potential is concomitantly shifted by the same amount, i.e. $\mu \mapsto \mu + V$. In either formulation, neither the charge distribution nor the charge-transfer variables minimize equations (7) and (17) respectively by this transformation. This then implies that the effective atomic electronegativities (30) are well-defined only up to a constant, and thus only electronegativity differences are physically meaningful in fluctuating-charge models.

Our results show that in QEq, the electronegativity of carbon and hydrogen in the four carbon systems do not change at all, which is at variance with our chemical intuition that carbon atoms in $sp^3$, $sp^2$ and sp hybridization environments ought to have different electronegativities. This is of course expected, as the intrinsic atomic electronegativities are constant parameters in fluctuating-charge models. In contrast, the effective atomic electronegativities in QTPIE show clear trends that have significant dependence on the hybridization environment. The electronegativities remain essentially unchanged between staggered and eclipsed ethane environments. However, the carbon electronegativities show a clear trend of increasing electronegativity C ($sp^3$) < C ($sp^2$) < C (sp), and the hydrogen electronegativities show a more ambiguous trend of decreasing electronegativity H ($sp^3$) ~ H ($sp^2$) < H (sp). The trend in the carbon electronegativities does indeed follow the expected trend from the Mulliken—Jaffé scale.

Despite the different trends in electronegativity between QEq and QTPIE, both charge models predict approximately the same trends in the charge distributions, with the charge on carbon exhibiting an increasing (less negative) trend C ($sp^3$) <

C (sp$^2$) < C (sp) and that on hydrogen showing a corresponding decreasing trend H (sp$^3$) > H (sp$^2$) > H (sp). These trends are reproduced in the Mulliken charge analysis, albeit with a smaller magnitude. However, the distributed monopole charges show very slight trends in the opposite sense.

Interestingly, QTPIE predicts almost the same charge distribution for ethane in the staggered and eclipsed conformers, as is the case for both *ab initio* charge analyses; whereas in contrast, the QEq charges vary by as much as 0.2 electron charges on the carbon between the two configurations. This reveals a previously unacknowledged advantage of the QTPIE model, that its charge distribution is relatively more stable with respect to conformational changes in this example. This property of the QTPIE model deserves further investigation.

## Conclusions

The preceding discussion reinforces the notion that empirical atomic electro­negativities must be environment dependent because the electron–accepting or electron–donating tendencies effects of atoms are never observed in isolation. The reason is obvious: there must be a counterpart to receive or donate charge. Even Pauling's seminal work on electronegativity and chemical bonding acknowledges that the electron–accepting or electron–donating tendencies of atoms depend on the other atoms in the molecule, and the best that can be hoped for is that this ten­dency for accepting or donating electrons be approximately constant over many different molecules, corresponding to many different chemical environments.

Despite this dependence on the chemical environment, electronegativities are indeed observed to vary only slightly depending on the exact molecule being con­sidered. This strongly suggests that there must be some underlying, intrinsic atomic property that is to a large extent responsible for the observed atomic elec­tronegativities. Here, we propose to call these quantities *bare* atomic electronega­tivities, as distinguished from the *effective* atomic electronegativities that are the only ones that can be directly observed.

Our work on developing fluctuating-charge models shows that indeed, physical constraints on the qualitative behavior of these models force us to use environ­ment-dependent electronegativities, and that their dependence on the chemical en­vironment is unavoidable and can be significant. However, this dependence is strongly local in the QTPIE model owing to the locality inherent in our definition of the distance attenuation factors. This locality is in accordance with our chemi­cal experiences. Furthermore, it is possible to reconcile this environment depend­ence with the existence of an intrinsic atomic quantity; in the QTPIE model, the "bare" atomic electronegativities that are intrinsic properties of individual atoms in isolation are distinguished from the *effective* atomic electronegativities that are system–specific.

## Appendix A. Fitting of *s*–type primitive Gaussian orbitals that best reproduce two–electron Coulomb integrals over *s*–type Slater orbitals

In this Appendix, we describe how we found the exponents given in **Table 1.**, which are of *s*–type primitive Gaussian orbitals such that the two–electron Coulomb integrals over them best reproduce those over *s*–type Slater orbitals. We construct these Gaussian orbitals by minimizing the norm of the $L^2$–difference between the homonuclear Coulomb integral over Slater orbitals and over Gaussian orbitals, i.e. given a Slater exponent $\zeta$, we want the Gaussian exponents $\alpha$ that minimizes

$$\left\| J^G(\alpha) - J^S(\zeta) \right\|_2^2 = \left\langle J^G(\alpha), J^G(\alpha) - 2J^S(\zeta) \right\rangle_2 + \left\| J^S(\zeta) \right\|_2^2 \quad (32)$$

where $\langle f, g \rangle_2 = \int\limits_0^\infty f(x)g(x)\,dx$ is the inner product in the function space $L^2[0, \infty)$, $\|f\|_2 = \sqrt{\langle f, f \rangle_2}$ is the $L^2$–norm, $J^G$ is the two–electron Coulomb integral over *s*–type primitive Gaussian orbitals

$$J^G(R; \alpha) = \frac{2\alpha}{\pi} \iint\limits_{\mathbb{R}^6} \frac{e^{-\alpha|r_1 - R|^2} e^{-\alpha|r_2|^2}}{|r_1 - r_2|}\,dr_1\,dr_2 = \frac{\mathrm{erf}\sqrt{\alpha}R}{R} \quad (33)$$

and $J^S$ is the two–electron Coulomb integral over *s*–type Slater orbitals

$$J^S(R; \zeta, n) = \frac{(2\zeta)^{4n+2}}{\left((2n)!\right)^2} \iint\limits_{\mathbb{R}^6} \frac{|r_1 - R|^n |r_2|^n e^{-\zeta|r_1 - R|} e^{-\zeta|r_2|}}{|r_1 - r_2|}\,dr_1\,dr_2 \quad (34)$$

which is given in closed–form in Ref. [49]. As the Slater exponent $\zeta$ is given for each minimization, the last term in ( 32) can be dropped without affecting the results of the minimization, and therefore the minimization problem is solved by the Gaussian exponent $\alpha$ that solves the equation

$$0 = \frac{\partial}{\partial \alpha} \left\langle J^G(\alpha), J^G(\alpha) - 2J^S(\zeta) \right\rangle_2 = \left\langle 2\frac{dJ^G(\alpha)}{d\alpha}, J^G(\alpha) - J^S(\zeta) \right\rangle_2$$

$$( 35)$$

We find the solution to ( 35) using the secant method with a trust radius of $\alpha/4$ at each iteration. The algorithm was terminated once the integral on the right hand side of ( 35) was less than $10^{-16}$ in absolute magnitude. The results are presented in Table 1., along with the maximum absolute error as defined by

$$\text{MAE} = \max_{0 \le R < \infty} \left| J^G\left(R; \alpha\right) - J^S\left(R; \zeta\right) \right| \qquad (36)$$

## Appendix B. Derivation of an explicit solution to fluctuating–charge models

In this Appendix, we derive the solution to the linear system of equations (9) that define a fluctuating-charge model, namely:

$$\begin{pmatrix} \mathbf{J} & \mathbf{1} \\ \mathbf{1}^T & 0 \end{pmatrix} \begin{pmatrix} \mathbf{q} \\ \mu \end{pmatrix} = \begin{pmatrix} -\boldsymbol{\chi} \\ Q \end{pmatrix} \qquad (37)$$

Assume $\mathbf{J}$ is invertible. As discussed in the main text, this should be true for all reasonable geometries without degenerate or nearly coincident atoms. Then we can perform Gaussian elimination on the second row by premultiplying the first row by $-\mathbf{1}^T \mathbf{J}^{-1}$ and adding the result to the second row. This transforms the system to

$$\begin{pmatrix} \mathbf{J} & \mathbf{1} \\ \mathbf{0}^T & 0 - \mathbf{1}^T \mathbf{J}^{-1} \mathbf{1} \end{pmatrix} \begin{pmatrix} \mathbf{q} \\ \mu \end{pmatrix} = \begin{pmatrix} -\boldsymbol{\chi} \\ Q + \mathbf{1}^T \mathbf{J}^{-1} \boldsymbol{\chi} \end{pmatrix} \qquad (38)$$

where $0 - \mathbf{1}^T \mathbf{J}^{-1} \mathbf{1}$ the Schur complement of $\mathbf{J}$ in this problem. Next, we solve the first row for $\mathbf{q}$, leading to

$$\begin{pmatrix} \mathbf{J} & \mathbf{0} \\ \mathbf{0}^T & -\mathbf{1}^T \mathbf{J}^{-1} \mathbf{1} \end{pmatrix} \begin{pmatrix} \mathbf{q} \\ \mu \end{pmatrix} = \begin{pmatrix} -\left(\boldsymbol{\chi} + \mathbf{1}\mu\right) \\ Q + \mathbf{1}^T \mathbf{J}^{-1} \boldsymbol{\chi} \end{pmatrix} \qquad (39)$$

This system of equations is now in block-diagonal form, and it is easy to write down the solution

$$\begin{pmatrix} \mathbf{q} \\ \mu \end{pmatrix} = \begin{pmatrix} -\mathbf{J}^{-1}\left(\boldsymbol{\chi} + \mathbf{1}\mu\right) \\ -\left(\mathbf{1}^T \mathbf{J}^{-1} \mathbf{1}\right)^{-1}\left(Q + \mathbf{1}^T \mathbf{J}^{-1} \boldsymbol{\chi}\right) \end{pmatrix} \qquad (40)$$

which is equivalent to (10). In particular, the explicit formula for the charge distribution is

$$\mathbf{q} = -\mathbf{J}^{-1}\boldsymbol{\chi} - \mu\mathbf{J}^{-1}\mathbf{1} = -\mathbf{J}^{-1}\boldsymbol{\chi} - \left(\frac{Q + \mathbf{1}^T\mathbf{J}^{-1}\boldsymbol{\chi}}{\mathbf{1}^T\mathbf{J}^{-1}\mathbf{1}}\right)\mathbf{J}^{-1}\mathbf{1} \qquad (41)$$

It is obvious that in general, the charge distribution that solves (9) is not $\mathbf{q} = -\mathbf{J}^{-1}\boldsymbol{\chi}$, which is the solution to the unconstrained problem $\mathbf{J}\mathbf{q} = -\boldsymbol{\chi}$, but contains an additional term that corresponds to a correction to account for the constraint of overall charge conservation. We belabor this point because to the best of our knowledge, the correct explicit solution (41) to a fluctuating-charge model has yet to appear in the literature.

Tables

| Element | Slater exponent[a] | Gaussian exponent | Error[b] |
|---------|--------------------|--------------------|----------|
| H | 1.0698 | 0.5434 | 0.01696 |
| Li | 0.4174 | 0.1668 | 0.00148 |
| C | 0.8563 | 0.2069 | 0.00162 |
| N | 0.9089 | 0.2214 | 0.00166 |
| O | 0.9745 | 0.2240 | 0.00167 |
| F | 0.9206 | 0.2313 | 0.00169 |
| Na | 0.4364 | 0.0959 | 0.00085 |
| Si | 0.7737 | 0.1052 | 0.00088 |
| P | 0.8257 | 0.1085 | 0.00089 |
| S | 0.8690 | 0.1156 | 0.00092 |
| Cl | 0.9154 | 0.1137 | 0.00091 |
| K | 0.4524 | 0.0602 | 0.00125 |
| Br | 1.0253 | 0.0701 | 0.00133 |
| Rb | 0.5162 | 0.0420 | 0.00121 |
| I | 1.0726 | 0.0686 | 0.00127 |
| Cs | 0.5663 | 0.0307 | 0.00114 |

[a] From Ref. [50].

[b] Maximum absolute error as defined in ( 36)

**Table 1.** Exponents of atomic orbital exponents that best reproduce the two–electron Slater integrals over the QEq orbitals.

| Property | Ethane (staggered) | Ethane (eclipsed) | Ethylene | Acetylene |
|---|---|---|---|---|
| Electronegativity (eV), $\chi_i$, in QEq(-H) | | | | |
| of C | 5.343 | 5.343 | 5.343 | 5.343 |
| of H | 4.528 | 4.528 | 4.528 | 4.528 |
| difference | 0.815 | 0.815 | 0.815 | 0.815 |
| Electronegativity (eV), $\widetilde{\chi}_i$, in QTPIE (30) | | | | |
| of C | −0.4251 | −0.4249 | −0.3358 | −0.1366 |
| of H | 0.2601 | 0.2593 | 0.2761 | 0.1907 |
| difference | 0.6852 | 0.6842 | 0.6119 | 0.3273 |
| Electronegativities (eV) on Mulliken-Jaffé scale[51] | | | | |
| of C | 8.15 | 8.15 | 8.91 | 10.42 |
| of H | 7.18 | 7.18 | 7.18 | 7.18 |
| difference | 0.97 | 0.97 | 1.73 | 3.24 |
| Atomic partial charge in QEq(-H) | | | | |
| of C | −1.0132 | −1.2660 | −0.5710 | −0.1052 |
| of H | 0.3377 | 0.4220 | 0.2855 | 0.1052 |
| Atomic partial charge in QTPIE | | | | |
| of C | −1.0644 | −1.0131 | −0.4287 | −0.0423 |
| of H | 0.3548 | 0.3377 | 0.2143 | 0.0423 |
| Mulliken charges from MP2/6-31G | | | | |
| of C | −0.4518 | −0.4646 | −0.3291 | −0.3254 |
| of H | 0.1506 | 0.1549 | 0.1645 | 0.3254 |
| Distributed monopole analysis from MP2/6-31G | | | | |
| of C | −0.0909 | −0.1185 | −0.1191 | −0.1569 |
| of H | 0.0303 | 0.0395 | 0.0596 | 0.1569 |

**Table 2**. Electronegativities and charge distributions of hydrocarbons with two carbon atoms calculated in QTPIE and QEq(-H), as well as electronegativities on the Mulliken-Jaffé scale [46] and charge analyses from MP2/6-31G calculations.